# Sunflower yield and oil quality interactions and variability: analysis through a simple simulation model

Gustavo A. Pereyra-Irujo, Luis A. N. Aguirrezábal *

*Unidad Integrada Balcarce, Facultad de Ciencias Agrarias - Universidad Nacional de Mar del Plata – Instituto Nacional de Tecnología Agropecuaria, CC 276, 7620 Balcarce, Argentina*

**Abstract**
Sunflower (*Helianthus annuus* L.) grain and oil quality are defined by grain weight and oil percentage, oil fatty acid composition and the amount of antioxidants. The aim of this work was to establish and validate a simple model, based on published relationships, which can estimate not only yield and its components, but also grain and oil quality aspects which are of relevance for industrial processes or human health. The model we developed provided good estimations of grain yield (similar to those of a more complex model) and oil quality from independent experiments. It explained known differences in potential yield and grain and oil quality between locations, in terms of differences in incident radiation, mean or minimum temperature. Simulations showed that recent climatic changes could have caused a decrease in sunflower yield and changes in oil quality. Our results suggest that at locations at lower latitudes, sunflower oil with high nutritious value and oxidative stability could compensate for relatively low yields, while at higher latitudes, high-linoleic acid oil production should be compatible with high yield potentials. Our model could facilitate the selection of the best location, sowing date or density for the production of sunflower oil with specific quality characteristics.
*Keywords: Sunflower; Crop model; Yield; Oil quality; Sowing date; Climate variability; Climate change*

\* *Corresponding author: Tel.: +54 2266 439100; fax: +54 2266 439101; E-mail: laguirre@mdp.edu.ar*

## 1. Introduction

The achievement of high grain and oil yields has been the main goal of sunflower production. In recent years, there has been an increasing demand for agricultural products with specific qualities (Siskos *et al.*, 2001). The suitability for alternative uses of sunflower oil is determined mainly by fatty acid composition and the amount of antioxidants. Oxidative stability, which delays the loss of nutritional value and the development of unpleasant flavors, depends on the proportion of oleic acid (18:1) and the amount of antioxidants, mainly α-tocopherol (Cabrini *et al.*, 2001). Alternatively, stability can be achieved by hydrogenation, although this process can produce trans fatty acids, which are considered to be atherogenic (Valenzuela and Morgado, 1999) and must be reported on food labels (Eller *et al.*, 2005). On the other hand, when oil methyl esther is used for fuel (biodiesel), a low degree of unsaturation is preferred (Neto da Silva *et al.*, 2003). From the point of view of human health, polyunsaturated acids, such as linoleic acid (18:2), are essential to mammals and have a potent hypocholesterolemic effect (Kris-Etherton and Yu, 1997), thus lowering the risk of cardiovascular disease. It has also been demonstrated that milk fat concentration of conjugated linoleic acid (CLA) –an anticarcinogen– from cows fed with high linoleic acid sunflower oil is about 500% greater that typical values (Kelly *et al.*, 1998). Additionally, tocopherols found in sunflower oil are precursors of vitamin E synthesis. Tocopherol requirements in humans are believed to depend on dietary content of polyunsaturated fatty acids (PUFA); a requirement of 0.6mg tocopherol per g PUFA has been suggested (American Heart Association, 2001).

Oil quality and yield are both dependent on the genotype, and its interaction with the environment. In specific cases, such as for high-oleic sunflower, the genotype can be the main determinant (Flagella *et al.*, 2002). For traditional sunflower genotypes, oil quality and yield depend largely on environment, which is highly variable among years, locations, and sowing dates within a single year. For crops with adequate water and nutrient supply, yield, grain quality and oil quality are determined primarily by solar radiation and temperature. The physiological bases of how these factors affect yield and oil quality have been extensively studied (Connor and Hall, 1997; Hall, 2004), and some experimental research has been done on how these factors can explain yield and oil quality variations between sowing dates, years and locations (*e.g.* de la Vega and Hall, 2002a, 2002b; Izquierdo *et al.*, 2006). Current knowledge has not been integrated so as to evaluate the variability and interactions between sunflower yield and quality.

Crop management is usually aimed at the achievement of high grain or oil yields. Agricultural practices or environmental conditions necessary for obtaining a specific oil composition may not necessarily result in high yields. Knowing if a specific oil composition is compatible with high yields, or whether low yields can be compensated for by the achievement of high quality grain or oil, could be valuable information for agronomic or commercial decision making. Establishing correlations between yield and quality for different environments (locations, sowing dates, etc.) through experimental research is a costly and time-consuming task.

Crop simulation models are a valuable tool for analyzing variability and complex interactions at a low cost and in a short time. Many sunflower models have been published that predict crop development and yield (Texier, 1992; Chapman *et al.*, 1993; Steer *et al.*, 1993; Villalobos *et al.*, 1996; Cabelguenne *et al.*, 1999; Stöckle *et al.*, 2003). Some of them can also simulate oil yield, but none of them predict oil quality. Rather than increasing the number of predicted variables, the tendency has been to strengthen the mechanistic bases of these models, allowing them to predict the effects of nitrogen or water stress, photoperiod, genotype, etc. For other crops, such as wheat (*Triticum aestivum* L.), models that predict grain quality have been established (*e.g.* Smith and Gooding, 1999), but they have not been used to establish relationships between yield and quality. Recently, a model has been developed for simulating yield and forage quality of maize (*Zea mays* L.; Herrmann *et al.*, 2005).

The quantitative nature of various studies on the response of yield components and oil quality to the environment (Cantagallo *et al.*, 1997; Aguirrezábal *et al.*, 2003; Sobrino *et al.*, 2003; Nolasco *et al.*, 2004; Izquierdo *et al.*, 2006) suggests the possibility of integrating them into a simple model that predicts yield and quality. Simple empirical models can be useful for rapidly analyzing the effects of sowing dates, locations, long-term climatic trends, etc. Moreover, a model that predicts yield as well as grain and oil quality could be used to analyze their interactions, and aid crop management in the selection of the best location or sowing date for obtaining a specific oil quality with highest yield.

The objectives of this work were: (i) to establish and validate a simple model to predict sunflower yield and oil quality based on weather data, (ii) to use the model to analyze the variability in yield and oil quality among locations, sowing dates, years and long-term climatic trends, and (iii) to explore the interactions between yield and oil quality.

## 2. Materials and methods

### 2.1. The model

The model was developed using published empirical relationships for estimating yield and its components (grain number and weight), grain quality (grain oil percentage) and oil quality (tocopherol concentration, oleic acid percentage, linoleic acid percentage and tocopherol : linoleic acid ratio), based on estimates of intercepted radiation, leaf area index and phenology. Figure 1 presents a flow chart that represents the model and the relationships between its variables. The model was established for *cv.* Dekalb G100. Some relationships established for other genotypes were recalibrated using experimental data. The model was programmed in a computer spreadsheet, for easy use and improvement.

#### 2.1.1. Phenology

Cummulative degree-days after sowing ($DD_s$, °Cd) are calculated from daily data for mean temperature ($T_m$, °C), as follows:

$$DD_s = \Sigma \; (T_m - T_b) \tag{1}$$

The model estimates the dates of occurrence of seven stages which are needed for model calculations (emergence, floral initiation, anthesis, beginning and end of the fatty-acid critical period, and beginning and end of the grain filling critical period) as the day in which DDs reaches a required value. Durations of phases between phenological stages were assumed to be controlled only by temperature, since *cv.* Dekalb G100 is considered to be insensitive to photoperiod (J. Re, Monsanto Argentina, personal communication). Although it has been demonstrated that base temperature ($T_b$, °C) differs between processes (*e.g.* -1°C for grain filling, Chimenti *et al.*, 2001), adequate phenology predictions can be obtained assuming a unique base temperature (*e.g.* Villalobos *et al.*, 1996; Aiken, 2005). A common base temperature of 6°C was used for all the phases, a value adopted by many authors (*e.g.* Kiniry *et al.*, 1992).

The required value used for the emergence stage was 100°Cd, which was based on those reported by Angus *et al.* (1981), Aguirrezábal *et al.* (1996) and Villalobos *et al.* (1996). Following the model of Villalobos et al (1996), and assuming no photoperiodic effect in *cv.* Dekalb G100, floral initiation is estimated to occur at approximately 1/3 of the emergence-anthesis phase. For this *cv.*, anthesis is estimated to require 900°Cd from emergence (value obtained from experiment 21, Table 1). The critical period for oil fatty acid determination occurs between 100 and 300°Cd after anthesis (Izquierdo *et al.*, 2006), while the critical period for grain filling occurs between 250 and 450°Cd after anthesis (Aguirrezábal *et al.*, 2003). Degree-days from emergence ($DD_e$, °Cd) and degree-days from anthesis ($DD_a$, °Cd) are calculated by subtracting 100°Cd or 900°Cd, respectively, from the DDs value.

#### 2.1.2. Leaf area

Total leaf area per plant (LAP, m$^2$) is estimated according to Sadras and Hall (1988), assuming leaf area dynamics to be controlled by mean temperature and plant density (PD, plants m$^{-2}$), and to depend on a maximum leaf area per plant ($LAP_{max}$, m$^2$), which is the value of LAP measured at a plant density of 1 pl m$^{-2}$. A LAP$_{max}$ value of 1.25 m$^2$ pl$^{-1}$ (V.R. Pereyra, unpublished data) was used for our calculations. Before anthesis, LAP is calculated as follows:

$$LAP = \exp(-5.68 + 0.0136 \cdot DD_e + 0.0000078 \cdot DD_e^2) \cdot \exp(0.2 - 0.16 \cdot PD) \cdot LAP_{max} \quad (2)$$

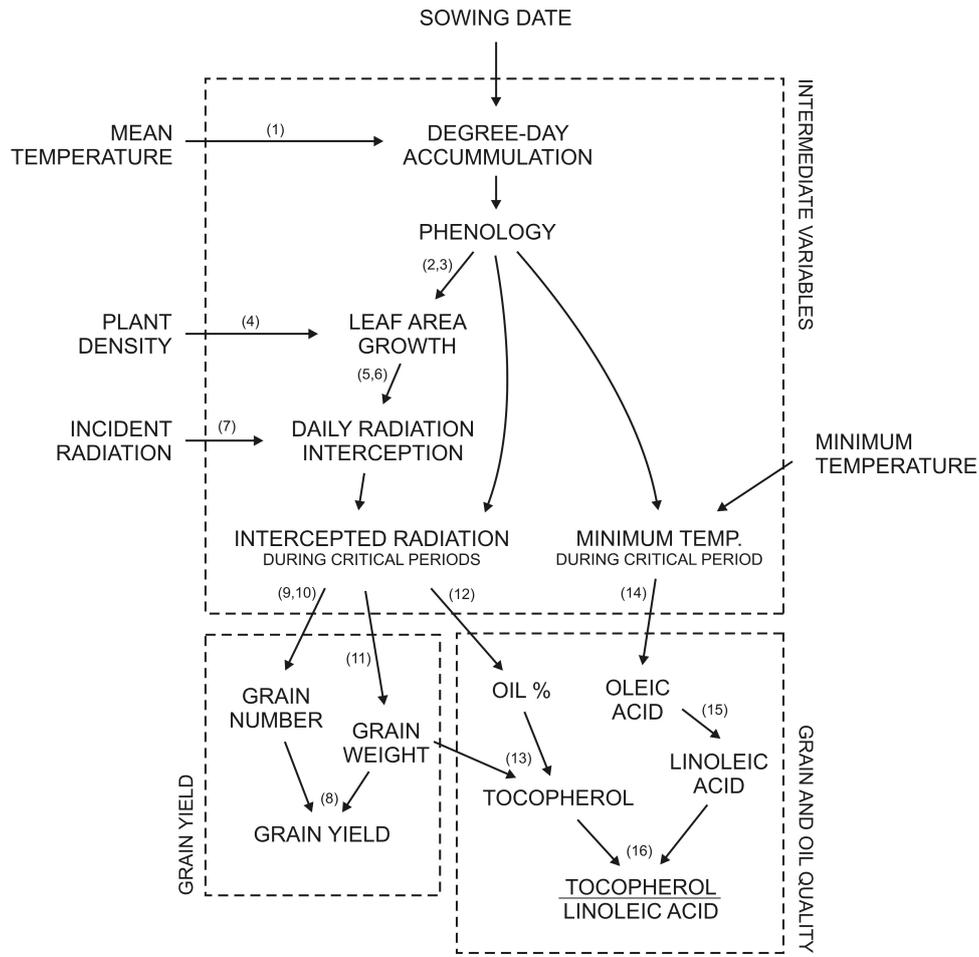

Figure 1. Schematic representation of the model, including input variables (outside the dashed lines), intermediate variables, output variables (grain yield, grain and oil quality) and their relationships. Numbers between paretheses indicate the equations in the 'Materials and Methods' section that represent each relationship.

The value of LAP reached at anthesis (LAP$_{anthesis}$, m$^2$) is maintained for 200°Cd, and then LAP is calculated as:

$$LAP = (-0.61 + 685/DD_a + 73870/DD_a^2) \cdot \exp(0.2 - 0.16 \cdot PD) \cdot LAP_{anthesis} \quad (3)$$

Leaf area index (LAI) is calculated as:

$$LAI = LAP \cdot PD \quad (4)$$

2.1.3. Radiation interception
The proportion of intercepted radiation at midday ($Q_{mid}$) is estimated according to the Lambert-Beer law (Gardner et al., 1985), with an extinction coefficient (k) value of 0.86 (Orgaz et al., 1992):

$$Q_{mid} = 1 - \exp(-k \cdot LAI) \quad (5)$$

This value is then corrected to account for the daily proportion of radiation interception ($Q_d$), according to Charles-Edwards and Lawn (1984):

$$Q_d = 2 \cdot Q_{mid} / (1 + Q_{mid}) \quad (6)$$

Daily intercepted radiation (IR, MJ m$^{-2}$ d$^{-1}$) is obtained from Qd and daily incident photosynthetically active radiation (PAR$_{inc}$, MJ m$^{-2}$ d$^{-1}$), which is calculated as 0.48 of daily incident global radiation (Bonhomme, 1993):

$$IR = Q_d \cdot PAR_{inc} \quad (7)$$

The sum of intercepted radiation is calculated for two different periods: (i) the period of grain number determination, between floral initiation and the beginning of seed filling (IRGN, MJ m$^{-2}$), and (ii) the grain filling critical period, between 250 and 450°Cd after flowering (IRGW, MJ m$^{-2}$).

2.1.4. Grain yield
Grain yield (GY, t ha$^{-1}$) is calculated as the product of its components, grain number (GN grains m$^{-2}$) and grain weight (GW, mg):

Table 1. Cultural details and seasonal weather conditions (from sowing to end of grain filling critical period) for experiments used for recalibration, verification or validation of the model.

| Exp. | Density (pl m$^{-2}$) | Sowing date | cv | Location | Seasonal weather conditions | | | | | Source † |
|---|---|---|---|---|---|---|---|---|---|---|
| | | | | | Daily max. temp. (°C) | Daily min. temp. (°C) | Daily PAR (MJ m$^{-2}$) | Precipit. (mm) | Potential evaporat. (mm) | |
| 1  | 4.0 | 20/09/1991 | Dekalb G100 | Balcarce | 22,8 | 11,4 | 8,9  | 497 | 521 | 1 |
| 2  | 7.0 | 20/09/1991 | Dekalb G100 | Balcarce | 22,8 | 11,4 | 8,9  | 497 | 521 | 1 |
| 3  | 3.8 | 29/10/1991 | Dekalb G100 | Balcarce | 24,9 | 13,3 | 9,2  | 471 | 482 | 1 |
| 4  | 6.8 | 29/10/1991 | Dekalb G100 | Balcarce | 24,9 | 13,3 | 9,2  | 471 | 482 | 1 |
| 5  | 3.6 | 29/11/1991 | Dekalb G100 | Balcarce | 26,2 | 14,5 | 9,0  | 442 | 438 | 1 |
| 6  | 7.3 | 29/11/1991 | Dekalb G100 | Balcarce | 26,2 | 14,5 | 9,0  | 442 | 438 | 1 |
| 7  | 4.1 | 30/12/1991 | Dekalb G100 | Balcarce | 24,5 | 13,5 | 7,8  | 438 | 384 | 1 |
| 8  | 7.0 | 30/12/1991 | Dekalb G100 | Balcarce | 24,5 | 13,5 | 7,8  | 438 | 384 | 1 |
| 9  | 1.0 | 20/10/1992 | Dekalb G100 | Balcarce | 24,7 | 12,0 | 9,6  | 262 | 525 | 2 |
| 10 | 2.1 | 20/10/1992 | Dekalb G100 | Balcarce | 24,7 | 12,0 | 9,6  | 262 | 525 | 2 |
| 11 | 4.1 | 20/10/1992 | Dekalb G100 | Balcarce | 24,7 | 12,0 | 9,6  | 262 | 525 | 2 |
| 12 | 1.0 | 20/11/1992 | Dekalb G100 | Balcarce | 26,3 | 13,2 | 9,6  | 180 | 493 | 2 |
| 13 | 2.1 | 20/11/1992 | Dekalb G100 | Balcarce | 26,3 | 13,2 | 9,6  | 180 | 493 | 2 |
| 14 | 4.1 | 20/11/1992 | Dekalb G100 | Balcarce | 26,3 | 13,2 | 9,6  | 180 | 493 | 2 |
| 15 | 1.0 | 16/12/1992 | Dekalb G100 | Balcarce | 27,0 | 13,7 | 8,8  | 208 | 455 | 2 |
| 16 | 2.1 | 16/12/1992 | Dekalb G100 | Balcarce | 27,0 | 13,7 | 8,8  | 208 | 455 | 2 |
| 17 | 4.1 | 16/12/1992 | Dekalb G100 | Balcarce | 27,0 | 13,7 | 8,8  | 208 | 455 | 2 |
| 18 | 5.8 | 17/11/1997 | Dekalb G100 | Balcarce | 24,1 | 12,0 | 8,9  | 312 | 459 | 3, 4 |
| 19 | 7.2 | 02/12/1999 | Dekalb G100 | Balcarce | 26,6 | 13,7 | 10,0 | 474 | 449 | 3 |
| 20 | 5.8 | 14/10/1992 | Dekalb G100 | Balcarce | 24,3 | 11,7 | 9,5  | 299 | 531 | 5 |
| 21 | 5.8 | 22/10/1993 | Dekalb G100 | Balcarce | 24,1 | 11,9 | 9,6  | 510 | 502 | 5 |
| 22 | 7.2 | 30/11/1993 | Dekalb G100 | Balcarce | 25,3 | 13,0 | 9,6  | 468 | 463 | 6 |
| 23 | 7.2 | 16/12/1994 | Dekalb G100 | Balcarce | 24,2 | 12,5 | 8,3  | 283 | 431 | 6 |
| 24 | 4.5 | 14/11/1995 | Dekalb G100 | Balcarce | 26,4 | 12,6 | 10,3 | 257 | 502 | 6 |
| 25 | 5.3 | 03/11/1998 | Dekasol 3881 | Balcarce | 26,7 | 12,3 | 10,0 | 129 | 505 | 7 |
| 26 | 7.3 | 22/11/1998 | Dekasol 3881 | Balcarce | 27,1 | 13,0 | 9,9  | 177 | 478 | 7 |
| 27 | 5.3 | 25/10/2004 | Dekasol 3881 | Balcarce | 25,1 | 13,1 | 10,0 | 252 | 490 | 8, 9 |
| 28 | 5.5 | 09/12/2004 | Dekasol 3881 | Paraná   | 29,7 | 18,1 | 11,3 | 376 | 488 | 8 |
| 29 | 6.1 | 17/11/1997 | Dekasol 3881 | Balcarce | 24,1 | 12,0 | 8,9  | 312 | 459 | 10 |
| 30 | 5.7 | 13/11/1991 | Dekalb G100 | Balcarce | 25,8 | 14,1 | 9,4  | 454 | 465 | 11 |

† Sources: 1. Andrade, F.H. (unpublished results); 2. Pereyra, V.R. (unpublished results); 3. Aguirrezábal et al. (2003); 4. Oil tocopherol data by Nolasco and Aguirrezábal (unpublished results); 5. Andrade and Ferreiro (1996); 6. Dosio et al. (2000); 7. Izquierdo et al. (2002); 8. Izquierdo et al. (2006); 9. Oil tocopherol data by Izquierdo and Aguirrezábal (unpublished results) ; 10. Nolasco et al. (2004); 11. Oliva and González (1996).

$$GY = GN \cdot GW \qquad (8)$$

Cantagallo et al. (1997) found no association between GN and GW in any of three hybrids grown at several locations and years, suggesting the lack of compensatory effects, at least for non-extreme source-sink ratios. Based on these experimental results, our model estimates these yield components through independent relationships.

A photothermal quotient (PQ, MJ m$^{-2}$ °C$^{-1}$) was used for the estimation of grain number (Cantagallo et al., 1997), which is is calculated from intercepted radiation (IRGN, MJ m$^{-2}$) and mean temperature (TGN, °C) between floral initiation and the beginning of seed filling, as follows:

$$PQ = IRGN / (TGN - T_b) \qquad (9)$$

Grain number is then estimated according to the following equation:

$$GN = 24997 \cdot PQ - 5919 \qquad (10)$$

This relationship was recalibrated for cv. Dekalb G100, using data from experiments conducted under optimum water and nutrient conditions (experiments 12-13-14-21, Table 1). Data used for recalibrations were not used for the validation of the model.

Grain weight was predicted according to Aguirrezábal et al. (2003), as a function of intercepted radiation between 250 and 450°Cd after anthesis (IRGW, MJ m$^{-2}$), with a maximum value dependent on plant density, as follows:

$$GW = \min \{ (22.70 + 1.05 \cdot IRGW / PD) , (81.60 - 4.75 \cdot PD) \} \qquad (11)$$

### 2.1.5. Grain and oil quality

Grain oil percentage (OP, %) is estimated according to Aguirrezábal et al. (2003), as a function of IRGW, with a maximum value set at 50%, as follows:

$$OP = \min \{ (36.4 + 0.5 \cdot IRGW / PD) , 50.0 \} \quad (12)$$

Tocopherol concentration (TC, µg·g$^{-1}$) is estimated as a function of oil weight per grain, according to Nolasco *et al.* (2004). Though this relationship was not established for *cv.* Dekalb G100, it showed to be valid for several sunflower hybrids.

$$TC = 563 + 1451 \cdot \exp(-0.1249 \cdot GW \cdot OP / 100) \quad (13)$$

Oleic acid percentage (OA, %) and linoleic acid percentage (LA, %) are estimated as a function of mean minimum temperature between 100 and 300°Cd after anthesis (T$_{min}$, °C), as follows:

$$OA = \min \{ [-23.1 + 3.4 \cdot (T_{min} + 0.9)] , 54.2 \} \quad (14)$$

$$LA = 84.575 - 0.938 \cdot OA \quad (15)$$

Fatty acid composition has been estimated as a function of night minimum temperature between 100 and 300°Cd after anthesis (Izquierdo *et al.*, 2006). Instead of night minimum temperature, which requires an hourly temperature data input, minimum temperature was used, for which data is usually more readily available. We determined that minimum temperature + 0.9°C did not significantly differ from night minimum temperature (p>0.4, r=0.99), with a root mean squared deviation (RMSD) lower that 0.5°C, using a set of hourly temperature data from a complete growing season (180 days) at Balcarce and Paraná. This correction performed similarly to an estimation of night minimum temperature from daily maximum and minimum temperature (Parton and Logan, 1981).

Though these relationships were established for *cv.* Dekasol 3881 (which was bred from a genetic pool related to that of *cv.* Dekalb G100, R. Reid, Monsanto Argentina, personal communication), they showed to be valid for different sunflower hybrids (Izquierdo *et al.*, 2006).

For the tocopherol : linoleic acid ratio (T:L, mg·g$^{-1}$), the amount of linoleic acid was calculated considering the unsaponifiable fraction of the oil. This fraction has been shown to be stable around a value of 0.9%, independently of oil concentration, fatty acid composition or genotype (N. Izquierdo and L. Aguirrezábal, personal communication). The equation used is the following (division by 10 for unit conversion):

$$T:L = TC / (LA \cdot 0.991) / 10 \quad (16)$$

*2.2. Validation*

First, a verification of the model was made by comparing the model estimations against data used for establishing or recalibrating the relationships. Second, a validation was performed by comparing output variables (grain number, grain weight, grain yield, oil percentage, fatty acid composition, tocopherol concentration) as well as intermediate variables (intercepted radiation during the grain number and grain weight critical periods, duration of sowing to emergence and emergence to anthesis phases), to independent experimental data, as suggested by Sinclair and Seligman (2000). Although all of the relationships used for the construction of the model had been previously validated, an evaluation of the overall performance of the model is important, especially in the cases where the final result depends on many intermediate steps (*e.g.* grain yield). For comparison purposes, we also evaluated estimations of phenology and grain yield made with OILCROP-SUN (Villalobos *et al.*, 1996), a more complex model whose mechanistic structure represents the current knowledge of sunflower physiology. Version 3.5 of this model was used, which includes the genetic coefficients for *cv.* Dekalb G100 (P1: 309, P2: 2.99, P5: 732, G2: 2800, G3: 1.81, O1: 65).

Model validation was performed by comparing model estimations to a set of independent data from the experiments described in Table 1. Data used for validation did not include data that had been used for recalibration or assembly of relationships (phenology data from experiments 21-25-26, grain number data from experiments 12-13-14-21, grain weight and oil percentage data from experiments 19-21-22-23-24, fatty acid composition data from experiments 25-26, and tocopherol concentration data from experiment 29). Grain yield data from experiments where either grain number or weight data had been used for recalibration or assembly of relationships, were not used for validation. Experiments used for phenology recalibration were not used for validation of any other variable.

All but one of the experiments were conducted at Balcarce, Argentina (37°45' S, 58°18' W) in a Typic Argiudoll soil (USDA Taxonomy). Experiment 28 was conducted at Paraná, Argentina (32°00' S, 60°15' W), in a Aquic Argiudoll soil (USDA Taxonomy). Most of the experiments were conducted with *cv.* Dekalb G100. Since insufficient oil-quality data was available for this *cv.*, data from the related *cv.* Dekasol 3881 was also used (experiments 25 to 29). In these cases, the requirement of degree-days from sowing to flowering was increased from 900 to 1040°Cd (value obtained from experiments 25 and 26, Table 1). Experiments 1 through 29 were performed under optimum water and nutrient conditions, and pests and diseases were adequately controlled. Data referred to as experiment 30 (Oliva and González, 1996) consist of data from a official comparative yield trial conducted at Balcarce, which were used only for validation of oleic and linoleic acid percentages (which are not usually affected by water or nutrient availability). Since not enough variation in oil tocopherol concentration could be obtained under non-limiting conditions, data from experiments manipulating the interception of incident radiation (shading and thinning treatments) were included. Treatments in experiments 18 and 29 are described in Aguirrezábal *et al.* (2003) and Nolasco *et al.* (2005), respectively. Treatments in experiment 27 were: (i) control, (ii) 50% shading, (iii) 80% shading, (iv) 38% thinning, and (v) 38% thinning + 50% shading.

Grain weight and number were determined by manually separating grains into non-empty and empty grains. Non-empty grains were counted, oven-dried (with air circulating at 60°C) and weighed. Grain weight was obtained by dividing the weight of all non-empty grains by the number of non-empty grains in each capitulum. Oil concentration was measured in duplicate samples by nuclear magnetic resonance and averaged. Grain weight and oil concentration are expressed on a dry weight basis. Fatty acid composition was determined by gas chromatography and expressed as a percentage of total fatty acid content as described by Izquierdo *et al.* (2002). Oil tocopherol concentration was determined by normal-phase HPLC as described by Nolasco *et al.* (2004).

Simulated and measured values were compared by two methods: (a) a mean squared deviation (MSD)-based analysis, and (b) a regression-based analysis.

The MSD-based method is the one proposed by Kobayashi and Salam (2000), in which the overall deviation of the model output is indicated by the root mean squared deviation (RMSD = $MSD^{1/2}$), and different aspects of this deviation are represented by the MSD components: the squared bias (SB), which represents the bias of the simulation from the measurement; the squared difference between standard deviations (SDSD), which measures the lack or excess of sensitivity of the model; and the lack of correlation weighted by the standard deviations (LCS), which indicates how the model fails to simulate the pattern of fluctuation across the measurements.

Additionally, a regression analysis was carried out, and the hypotheses of intercept = 0 and slope = 1 were tested. The correlation coefficients (r) were calculated in each case.

### 2.3. Simulations

Simulations were performed using a set of weather data for the years 1971–2005 from three locations in Argentina: Balcarce, H. Ascasubi (39°22′ S, 62°39′ W) and Paraná. These locations have contrasting radiation and temperature regimes (Table 2). The model output variables were estimated for each year, for a range of sowing dates which started on the first day with mean temperature ≥ 14°C (Table 2) and ended on December 31st. Plant densities evaluated were 3.0, 4.5 and 6.0 pl·m$^{-2}$.

In order to estimate the effect of sowing date on yield and quality, and to assess the variability between locations, mean values (years 1971–2005) were calculated for each sowing date and each location.

The occurrence of long-term climatic trends in input variables (temperature and PAR) and their effect on output variables (phenology, yield and its components, and oil quality variables) were evaluated. The significance of these trends was tested by the Spearman rank correlation test. This nonparametric method is free of the assumption that the data being analyzed have normal distributions with equal variances and so does not emphasize extreme values. The change in input or output variables was calculated as the the difference between estimated values for 2005 and 1971, and expressed as a percentage of the estimated value for 1971.

Correlation analyses were performed to assess the possible interactions between oil quality variables and grain yield. These were performed either on a pool of all simulation results (all years, sowing dates, locations and plant densities, n > 10000), or separated by location or plant density.

### 3. Results

#### 3.1. Validation

The model estimations verified experimental data used for establishing or recalibrating the relationships (Figure 2, empty symbols). When compared to data from independent experiments, the model predicted the overall trends in all the analyzed variables (Figure 2, filled symbols). Measured values for fatty-acid composition and tocopherol concentration were well predicted irrespective of the *cv.* used (Dekasol 3881 or Dekalb G100). For nearly all evaluated variables, LCS was the major component of the MSD (Table 3), indicating that the model had neither a significant bias of the simulation from the measurement (low SB) nor a lack or excess of sensitivity (low SDSD). The intercept was not found to be statistically different from 0 in any case. The slope was found to be significantly (p=0.035) different from 1 only for grain weight estimations. Nevertheless, when we tested the hypothesis that intercept and slope are simultaneously different from 0 and 1, respectively, differences were not significant in any case (the p value obtained for grain weight was 0.16).

Table 2
Mean (1971-2005) daily incident PAR and daily mean temperature during the sunflower growing season (September–March), and earliest sowing date used in the simulations (first day with mean temperature ≥ 14°C) for Balcarce, H. Ascasubi and Paraná.

|  | Daily incident PAR (MJ·m$^{-2}$) | Daily mean temp. (°C) | Earliest sowing date |
|---|---|---|---|
| Balcarce | 8.6 | 16.5 | October 20th |
| H. Ascasubi | 9.9 | 17.1 | October 12th |
| Paraná | 9.8 | 21.4 | September 1st |

The more complex model OILCROP-SUN (Villalobos *et al.*, 1996) did not show an improved estimation of phenology or grain yield, when tested against the same validation data set (compare Tables 3 and 4). Particularly, the estimations of the duration of the emergence-anthesis phase showed a higher RMSD (5.1 days, compared to 2.3 days in our model).

*3.2. Simulations*

Estimated yields for a given sowing date, averaged over the 1971-2005 period, were consistently highest at H. Ascasubi, intermediate at Balcarce and lowest at Paraná (Figure 3a). Highest yields at H. Ascasubi were due to a combination of relatively low temperature and high incident radiation. Balcarce showed a slightly lower temperature, but a much lower incident radiation which accounted for the lower yield potential; lowest estimated yields at Paraná were due to a much higher temperature, despite an incident radiation similar to that of H. Ascasubi (Table 2).

Yields tended to decrease when sowing date was delayed, at all three locations (Figure 3a). This was due to the fact that, on average, temperature increases more rapidly than radiation before the summer solstice, and radiation decreases more rapidly than temperature afterwards, causing a steady decline in the photothermal quotient. Grain oil percentage showed a slight decrease associated with the delay of sowing, but with a different magnitude according to the location. At H. Ascasubi, grain oil percentage was reduced by less than 1 percentage point a month, while at Balcarce, a one-month delay resulted in a reduction in 2 percentage points (results not shown). This was due to the fact that at H. Ascasubi estimated grain oil percentage at early sowings is close to the maximum in most of the years. The estimated response of fatty acid composition to sowing date was not the same at all locations. At Paraná, oleic acid percentage increased (Figure 3b) and linoleic acid percentage decreased (results not shown) when sowing date was delayed, while at Balcarce and H. Ascasubi estimates of fatty acid composition remained almost constant within the season (Figure 3b). Estimated oil weight per seed and oil tocopherol concentration showed small variation among locations, years and sowing dates, but showing a slight effect of sowing density. Variations in the tocopherol : linoleic acid ratio were therefore determined mostly by variations in fatty acid composition. At Balcarce and H. Ascasubi, this ratio remained fairly stable across years and sowing dates, around a value of 1 mg·g$^{-1}$. At Paraná, values of nearly 2 mg·g$^{-1}$ could be reached with late sowings. In all cases, these values were slightly increased at higher sowing densities, due to a decreased oil weight per grain, which reduces the dilution effect described by Nolasco *et al.* (2004).

General tendencies were reversed in some years, for which the model predicted higher yields at Balcarce than at H. Ascasubi (e.g year 1979: maximum estimated yield of about 5300 kg·ha$^{-1}$ at Balcarce vs.

Figure 2. Estimated vs. observed data for phenology (a), grain number (b), grain weight (c), grain yield (d), grain oil percentage (e), oleic acid percentage (f, triangles), linoleic acid percentage (f, squares), and tocopherol concentration (g). Filled symbols correspond to independent data, which were used for validation analyses. Empty symbols correspond to data used to establish or recalibrate the relationships. Dotted lines are regression lines (slope and intercept values are given in Table 3) between estimated results and independent data. Solid lines represent the 1:1 line.

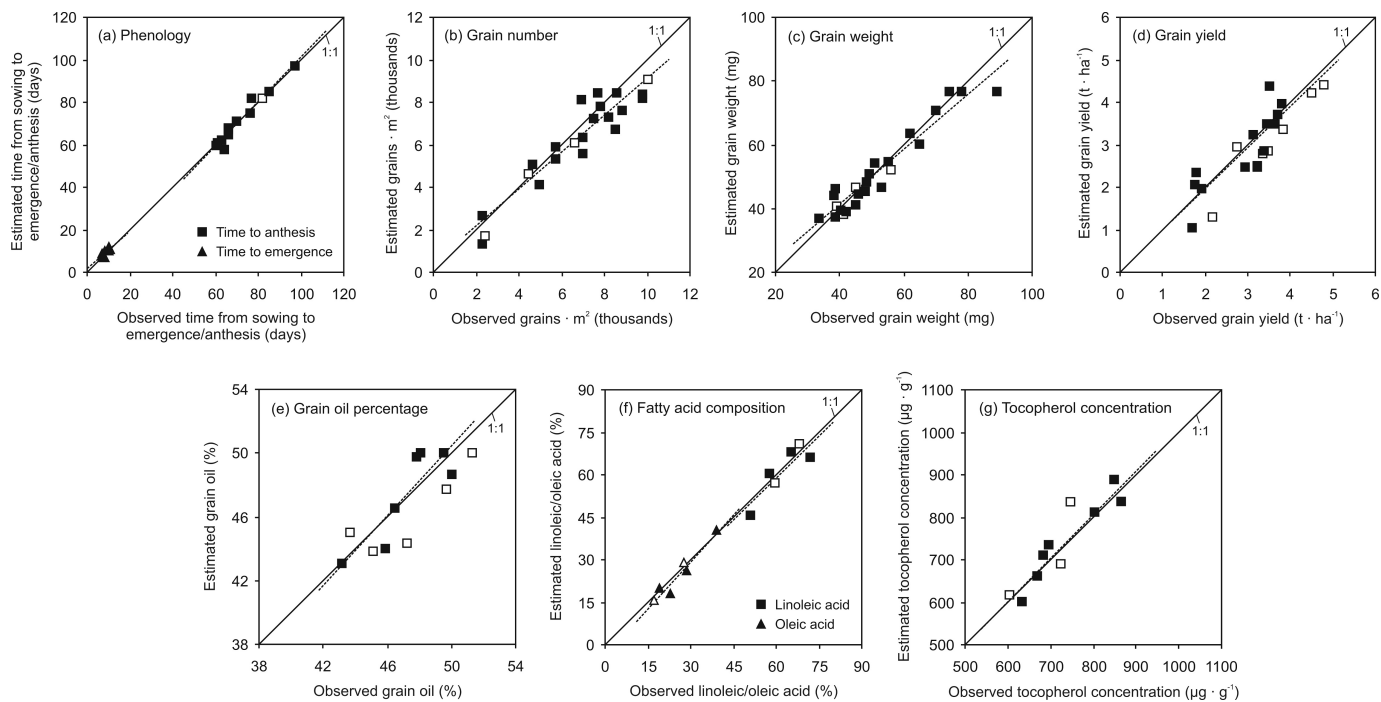

Table 3
Results of the Mean Squared Deviation (MSD)-based and Regression-based analyses of the differences between simulated and measured values. SB, SDSD and LCS expressed as percentage of MSD. Significance of the differences between the regression line and the 1:1 line (intercept = 0, slope = 1) are indicated (*: p<0.05, ns: p>0.05).

| Variable | MSD-based analysis | | | | Regression-based analysis | | |
|---|---|---|---|---|---|---|---|
| | RMSD | SB | SDSD | LCS | Intercept | Slope | Correlation coefficient |
| Sowing-emergence | 1.3 d | 36% | 2% | 62% | 1.7 ns | 0.88 ns | 0.79 |
| Emergence-anthesis | 2.3 d | 0% | 10% | 90% | -3.4 ns | 1.04 ns | 0.98 |
| Grain number | 952 grains $\cdot m^{-2}$ | 23% | 3% | 75% | 532 ns | 0.86 ns | 0.92 |
| Grain weight | 4.4 mg | 1% | 12% | 87% | 6.9 ns | 0.86 * | 0.96 |
| Grain yield | 0.48 t $\cdot ha^{-1}$ | 3% | 5% | 92% | 30 ns | 0.96 ns | 0.84 |
| Oil percentage | 1.4 % | 1% | 16% | 82% | -4.1 ns | 1.09 ns | 0.87 |
| Oleic acid percentage | 2.8 % | 16% | 17% | 68% | -4 ns | 1.11 ns | 0.96 |
| Linoleic acid percentage | 4.9 % | 2% | 5% | 93% | -0.3 ns | 0.98 ns | 0.88 |
| Tocopherol concentration | 20 $\mu g \cdot g^{-1}$ | 8% | 5% | 87% | -12 ns | 1.03 ns | 0.96 |
| IRGN | 3.9 MJ | 27% | 2% | 70% | 1.3 ns | 0.99 ns | 0.99 |
| IRGW | 3.9 MJ | 9% | 2% | 90% | 1.2 ns | 0.92 ns | 0.85 |

RMSD: Root mean squared deviation
SB: Squared bias
SDSD: Squared difference between standard deviations
LCS: Lack of correlation weighted by the standard deviations
IRGN: sum of intercepted radiation per plant between floral initiation and 250°Cd after anthesis
IRGW: sum of intercepted radiation per plant between 250 and 450°Cd after anthesis

4500 kg $\cdot ha^{-1}$ at H. Ascasubi) or an increase in yields of late sowings (*e.g.* year 1982 at Balcarce: an increase of 354 kg $\cdot ha^{-1}$ per month vs. a decrease of 613 kg $\cdot ha^{-1}$ per month for the 1971-2005 period). Also, in some years, the estimated oleic acid percentage at Balcarce was as high as at Paraná (*e.g.* year 1991: values of about 38% at both locations in early November sowings). Despite the apparent insensitivity of fatty acid composition to sowing date at Balcarce, for some years the model estimates an increase (*e.g.* year 1979: 5.8% per month) or a decrease (*e.g.* year 1982: -1.2% per month) in oleic acid percentage when sowing is delayed. In general, oleic acid percentage tended to increase with delayed sowings in years in which estimated potential yield decreased the most, whereas it was estimated to decrease in years that yield showed almost no change, or even increased with delayed sowing dates (results not shown). While at Balcarce and H. Ascasubi increased oleic acid of late sowings occurred in about 30% or 60% of the years, respectively, at Paraná oleic acid percentage was estimated to increase in all but one year.

A significant decreasing trend in estimated yield during the 1971-2005 years was detected at Balcarce (-21% over the 35-year period, p<0.01, Figure 4a) and H. Ascasubi (-13%, p<0.05), but not at Paraná (-12%, p>0.05). The estimated reduction in potential yield was higher at Balcarce than at H. Ascasubi, due to the greater magnitude of the changes in incident PAR and temperature (Table 5). A significant reduction in incident PAR was observed also at Paraná, but this did not result in a significant change in potential yield. Estimated reductions in oil yield were slightly higher than for grain yield due to an additional reduction in grain oil percentage (results not shown). Significant increasing or decreasing trends during the

Table 4
Results of the Mean Squared Deviation (MSD)-based and Regression-based analyses of the differences between measured values (same as those used in Table 3) and values simulated with OILCROP-SUN. SB, SDSD and LCS expressed as percentage of MSD. Significance of the differences between the regression line and the 1:1 line (intercept = 0, slope = 1) are indicated (*: p<0.05, **: p<0.01).

| Variable | MSD-based analysis | | | | Regression-based analysis | | |
|---|---|---|---|---|---|---|---|
| | RMSD | SB | SDSD | LCS | Intercept | Slope | Correlation coefficient |
| Sowing-emergence | 1.7 d | 33% | 4% | 63% | 6.4 * | 0.35 * | 0.45 |
| Emergence-anthesis | 5.1 d | 77% | 10% | 13% | 17.1 ** | 0.82 ** | 0.98 |
| Grain yield | 0.45 t $\cdot ha^{-1}$ | 8% | 0% | 92% | 338 * | 0.84 * | 0.84 |

RMSD: Root mean squared deviation
SB: Squared bias
SDSD: Squared difference between standard deviations
LCS: Lack of correlation weighted by the standard deviations

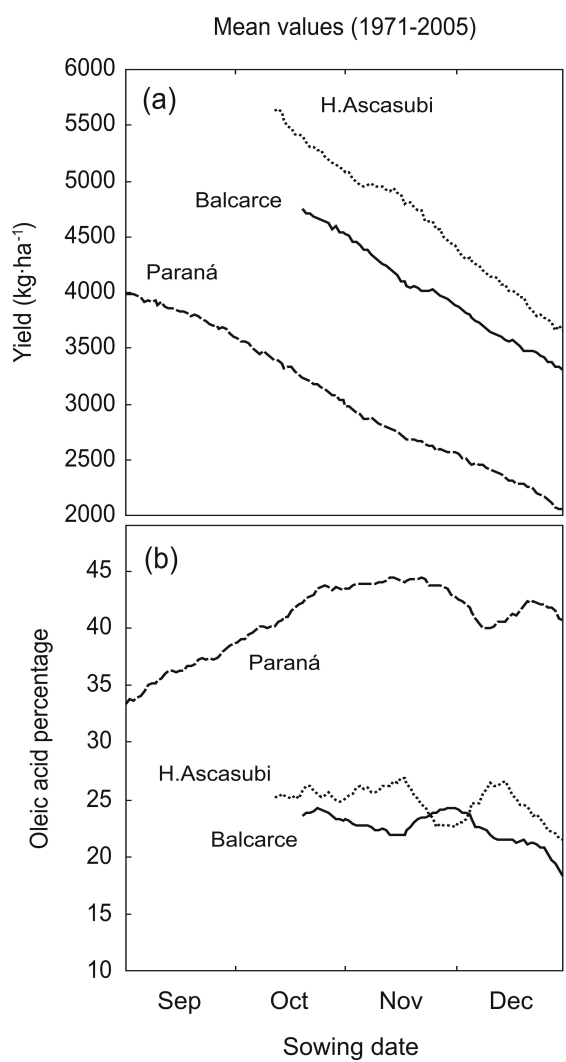
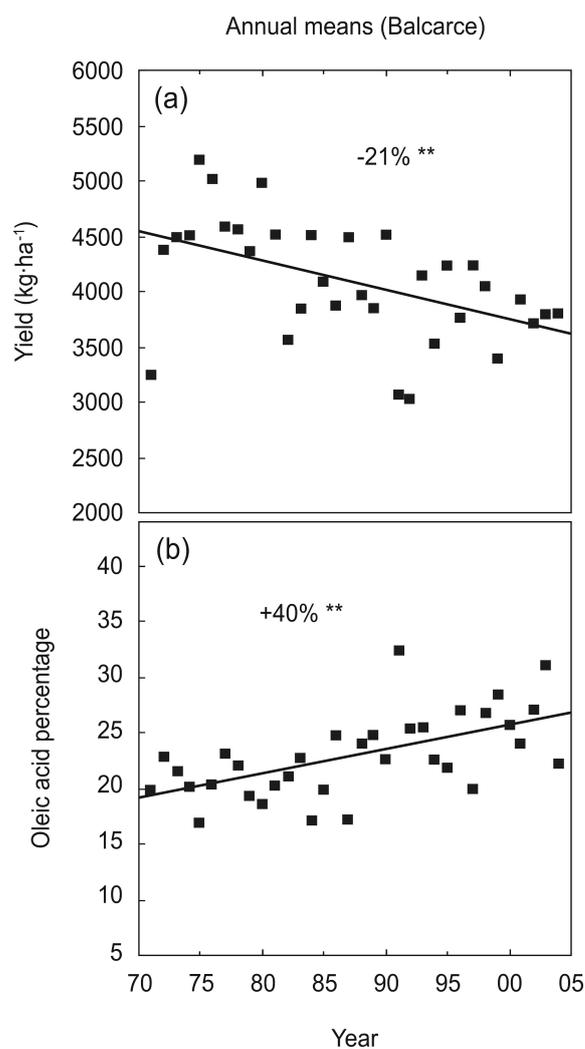

Figure 3. (a) Estimated yield, averaged over the 1971-2005 period, for each sowing date and location. (b) Estimated oleic acid percentage, averaged over the 1971-2005 period, for each sowing date and location.

Figure 4. (a) Annual means of estimated yield, from 1971 through 2005 at Balcarce. (b) Annual means of estimated oleic acid percentage, from 1971 through 2005. Solid lines are regression lines. Percentage change (difference between regression values for 2005 and 1971) and significance of Spearman rank correlation test are indicated (**: p<0.01).

1971-2005 period were found for all grain and oil quality variables at Balcarce and H. Ascasubi, of greater magnitude at the former location (*e.g.* oleic acid percentage, Figure 4b). Changes in grain and oil quality were not only due to changes in incident PAR and mean temperature, but also due to a greater increase in minimum temperature (Table 5).

At Balcarce, the significant increase in temperature during the 1971-2005 period (Table 5) lead to an average 7-day reduction in estimated time from sowing to flowering (results not shown, p<0.01). Thus, in recent years, the critical periods for yield components tended to occur nearer the summer solstice, with higher probability of better radiation and temperature conditions than in the 1970's. By regressing simulated yield vs. flowering date, the effect of a 7-day advance in flowering time between 1971 and 2005 can be estimated to equal 180 kg·ha$^{-1}$ (4%). This indirect, positive effect of increased temperature partly compensates for the direct, negative effect of increased temperature and decreased incident radiation on yield that can be deduced from equations 9 to 11. Had phenology not been affected, the effect of climate change on yield at Balcarce would have been larger. Also because of this significant increase in temperature at Balcarce, the day in which temperature is ≥ 14°C tended to occur 5 days earlier in 2005 than in 1971. As can be calculated from results shown in Figure 3a, this 5-day advance in sowing date could provide a 100 kg·ha$^{-1}$ (2%) increase in potential yield. This implies that, if sowing date had not been held constant in our simulations, the estimations of the effect of climate change on yield at Balcarce would have been slightly smaller. Regarding oil composition, since no effect of sowing date was observed at this location, changes in phenology or in the first possible sowing date would not be able to compensate for the effects of the increase in minimum temperature during the 1971-2005 period.

Grain oil content was generally positively correlated with yield (r=0.71). This correlation can be explained by the fact that critical periods for oil content and grain weight are highly coincidental (Aguirrezábal *et al.*, 2003). Linoleic acid was positively correlated (r=0.66), and oleic acid negatively correlated (r=-0.66), with yield. This correlation found in the pooled results was mainly due to temperature differences between locations. When considering one location at a time, the highest correlation between oil composition and yield was observed at Paraná (r=0.53), while it was weaker at Balcarce and H. Ascasubi (r=0.33 at both locations). Oil tocopherol concentration was negatively correlated with yield (r=-0.62). In this case, the response was the same at all locations, but was found to depend strongly on sowing density. At a density of 3.0 pl·m$^{-2}$ most estimated values fell between 600 and 650 µg·g$^{-1}$, while at a density of 6.0 pl·m$^{-2}$ estimated values ranged between 620 and 825 µg·g$^{-1}$ and were negatively correlated with yield (r=-0.62). This was because at higher densities, the same yield can be obtained with a higher number of smaller grains, for which the response curve of tocopherol concentration to variations in oil content is steepest.

Table 5
Change (difference between regression values for 2005 and 1971) in incident PAR, mean temperature and minimum temperature for the sunflower growing season (September–March), during the 1971-2005 period at the three studied locations. Significance of Spearman rank correlation test is indicated in each case (*: p<0.05, **: p<0.01, ns: p>0.05).

|  | Incident PAR (MJ·m$^{-2}$·d$^{-1}$) | Mean temp. (°C) | Min. temp. (°C) |
|---|---|---|---|
| Balcarce | -0.7 ** | +1.5 ** | +2.0 ** |
| H. Ascasubi | -0.4 * | +0.9 * | +1.2 ** |
| Paraná | -0.7 ** | +0.4 ns | +0.3 ns |

## 4. Discussion

The model was capable of predicting yield and oil quality of sunflower under non-limiting conditions. For sunflower, this is the first model capable of predicting oil quality as well as grain yield. Estimations of our model agreed adequately with experimental data over a wide range of environmental conditions. Although validation data belongs mainly to one location, there was a large variation in environmental factors due to year or sowing date (Table 1). As an example, the ratio between maximum and minimum yields in the validation data set was 2.3:1, whereas the ratio between estimated yields at Ascasubi and Paraná was 1.7:1. This allows us to presume the validity of the model for other locations as well.

A good prediction of intermediate variables indicates the general consistency of the model (Sinclair and Seligman, 2000). Despite the accumulation of errors derived from the integration of empirical relationships from different sources, the model provided good estimation of output variables, as indicated by the relatively low RMSD values. A good estimation not only of mean values, but also of the variability in the experimental data was evidenced by the major contribution of the LCS component to the MSD. Our relatively simple model showed a performance similar to that of OILCROP-SUN. Besides, our model was capable of predicting variables not considered by current sunflower crop models. We suggest that our model is adequate for analyzing yield–oil quality interactions and variability among different locations, years and sowing dates, and to evaluate the impact of climate change (Carbone *et al.*, 2003).

While the performance of our model was good under non-limiting conditions, it does not consider effects triggered by water or nutrient stresses, as other more complex models do. Depending on timing and intensity, water stress could decrease yield directly by affecting grain set or grain filling rate, or indirectly by decreasing leaf expansion and photosynthesis, or increasing leaf senescence (Connor and Hall, 1997; Hall, 2004). Stress effects could be incorporated into our model by modifying different relationships and including water or nutrient balance calculations, but this would increase its complexity, which is against the objectives of our work.

The estimated grain and oil yield response to sowing date under non-limiting conditions was consistent with many published studies that show decreased yield when sowing is delayed (de la Vega and Hall, 2002a, and references therein). Differences between locations in estimated grain and oil yield are consistent with experimental data, which show that northern locations, such as Paraná, reach lower yields than central or southern locations (Chapman and de la Vega, 2002) and that H. Ascasubi has a higher yield potential that Balcarce (Sadras, 2003). In the studied cases, differences between locations were either due to temperature (Paraná) or incident radiation (H. Ascasubi). The estimated variability between years in potential yield and in its response to sowing date could be a likely explanation for the well-known strong seasonal variation of sunflower yield in Argentina (Chapman and de la Vega, 2002), since our results show that the combined effects of incident radiation and temperature account for quantitatively similar variations in yield as those shown in their experiments.

Estimations of oil fatty acid composition agreed with the range of values reported by Muratorio *et al.* (2003) for Balcarce (18-32% oleic acid) and Paraná (22-50% oleic acid). Oil compositions estimated for late sowings at Paraná were close to those of NuSun (mid-oleic) sunflower oil (57% oleic acid, 32% linoleic acid), which provides a balance of unsaturated fatty acids near the optimum for lowering the risk of cardiovascular disease (Binkoski *et al.*, 2005). Moreover, oleic acid values of 58% have been reported for traditional sunflower grown at even lower latitudes (Muratorio *et al.*, 2003).

The range of estimated tocopherol concentrations agreed with the range of values observed by Nolasco *et al.* (2004) in their control treatments, grown under non-limiting conditions (approximately 550-750 µg·g$^{-1}$). The fact that any stress affecting final grain weight or oil concentration could increase tocopherol concentration would explain higher values reported in the literature, which typically reach 900 µg·g$^{-1}$ (Gunstone *et al.*, 1994) with extreme values of up to 1973 µg·g$^{-1}$ (Velasco *et al.*, 2002). At all locations, years and sowing dates, estimated tocopherol : linoleic acid ratio exceeded the minimum recommended value of 0.6 mg·g$^{-1}$ (American Heart Association, 2001). However, considering that a fraction of oil tocopherol is usually lost during refinement and storage (Cert *et al.*, 2000), values significantly higher than this lower limit should be sought.

Significant reductions in potential yield and changes in oil quality during the 1971-2005 period were estimated for some locations. These estimations derived from climatic trends which coincide with global phenomena, *i.e.* global warming (IPCC, 2001) and global dimming (Stanhill and Cohen, 2001). Besides direct effects of changes in incident radiation and temperature that could have been predicted with individual relationships, simulations revealed more complex responses involving changes in phenology and changes in the beginning of the growing season, which in turn could affect yield and oil quality. It should be noted, however, that these effects could be reduced in photoperiod-sensitive genotypes if changes in flowering date are smaller than those expected by changes in temperature alone. Changes in the beginning of the growing season following the raise in air temperature, similar to those estimated here, have been observed during the 1969-1998 period in Europe through the unfolding of leaves in four tree species (Chmielewski and Rötzer, 2001).

Our results suggest the possibility of selecting the best location, sowing date and sowing density for the production of a specific grain or oil quality with highest yield. At locations at higher latitudes, such as Balcarce and Ascasubi, oil with relatively high linoleic acid content (70-90%, for margarine production, dairy cow feed or human consumption) could be obtained almost regardless of sowing date or density. Furthermore, this oil quality should be compatible with the achievement of high grain yields. On the other hand, at locations at lower latitudes, such as Paraná, sunflower oil with high oxidative stability and high nutritious value (close to mid-oleic fatty acid composition, tocopherol : linoleic acid ratio > 2 mg·g$^{-1}$) could be obtained through late sowings and relatively high densities. This oil quality could compensate for the lower yield potential of late sowings. Early sowing of mid- or high-oleic cultivars would be a way to avoid this negative correlation.

## 5. Conclusions

A simple model based on published relationships, which can estimate not only yield and its components, but also grain and oil quality aspects which are of relevance for industrial processes or human health was established and validated. It explained differences in potential yield and grain and oil quality between locations, in terms of differences in incident radiation, mean or minimum temperature. Analyses of the interactions between yield and oil quality suggest that at low latitude locations, sunflower oil with high nutritious value and oxidative stability could compensate for relatively low yields, while at higher latitudes, high-linoleic acid oil production should be compatible with high yield potentials.

### Acknowledgements

We are grateful to Fernando H. Andrade, Natalia Izquierdo, Susana Nolasco and Víctor R. Pereyra for kindly allowing the utilization of their unpublished experimental data, and to Octavio Caviglia, Jorge Cepeda and Oscar Valentinuz for providing weather data. We also thank Fernando H. Andrade for carefully revising the manuscript. Luis Aguirrezábal is a member of CONICET. Gustavo Pereyra-Irujo received scholarships from UNMdP and CONICET. This work was supported by Agencia Nacional de Promoción Científica y Tecnológica (PICT 08-09711), Asociación Argentina de Girasol (Proyectos 2002), Universidad Nacional de Mar del Plata and Instituto Nacional de Tecnología Agropecuaria.